\documentclass[iop]{emulateapj}

\usepackage{hyperref}
\usepackage[bottom]{footmisc}

\shorttitle{electrostatic turbulence in collisionless shocks}
\shortauthors{Wang et al.}

\begin{document}


\title{electrostatic turbulence and Debye-scale structures in collisionless shocks}


\author{R. Wang\altaffilmark{1,2}, I.Y. Vasko\altaffilmark{1}, F.S. Mozer\altaffilmark{1}, S.D. Bale\altaffilmark{1, 2}, A.V. Artemyev\altaffilmark{4,5}, J.W. Bonnell\altaffilmark{1},  R. Ergun\altaffilmark{6}, B. Giles\altaffilmark{7}, P.-A. Lindqvist\altaffilmark{8}, C.T. Russell\altaffilmark{5} and R. Strangeway\altaffilmark{5}}
\email{rachel\_w@berkeley.edu }



\altaffiltext{1}{Space Sciences Laboratory, University of California, Berkeley, CA 94720}
\altaffiltext{2}{Physics Department, University of California, Berkeley, CA 94720}
\altaffiltext{4}{Space Research Institute of Russian Academy of Sciences, Moscow, Russia}
\altaffiltext{5}{Institute of Geophysics and Planetary Sciences, University of California, Los Angeles, USA}
\altaffiltext{6}{University of Colorado at Boulder, Boulder, CO, USA}
\altaffiltext{7}{NASA, Goddard Space Flight Center, Greenbelt, Maryland, USA} \altaffiltext{8}{Royal Institute of Technology,
Stockholm, Sweden}


\begin{abstract}
We present analysis of more than one hundred large-amplitude bipolar electrostatic structures in a quasi-perpendicular supercritical Earth's bow shock crossing, measured by the Magnetospheric Multiscale spacecraft. The occurrence of the bipolar structures is shown to be tightly correlated with magnetic field gradients in the shock transition region. The bipolar structures have negative electrostatic potentials and spatial scales of a few Debye lengths. The bipolar structures propagate highly oblique to the shock normal with velocities (in the plasma rest frame) of the order of the ion-acoustic velocity. We argue that the bipolar structures are ion phase space holes produced by the two-stream instability between incoming and reflected ions. This is the first identification of the ion two-stream instability in collisionless shocks. The implications for electron acceleration are discussed.
\end{abstract}





\section{Introduction}


Supercritical quasi-perpendicular shocks are of interest because of relatively efficient electron acceleration in the shock transition region as inferred from observations in the Earth's bow shock \citep[][]{Gosling89,Oka06} and astrophysical shocks \citep[e.g.,][]{Bamba03,vanWeeren10:sci}. In supercritical quasi-perpendicular shocks, the reflection of a fraction of incoming ions \citep[e.g.,][]{Leroy82} gives rise to various wave activities potentially involved in electron acceleration \citep[e.g.,][]{Papadopoulos85}. Numerical simulations demonstrated that, at high Mach numbers, electrostatic turbulence driven by the Buneman instability may provide efficient electron acceleration in the shock transition region \citep[e.g.,][]{cargill88,Hoshino02,Schmitz02b,Shimada&Hoshino04,Amano09}. Similar process of electron acceleration by electrostatic turbulence may operate at lower Mach numbers typical in the Earth's bow shock \citep[e.g., simulations by][]{Umeda09}. Nevertheless, the lack of detailed experimental analysis of the origin of electrostatic turbulence in collisionless shocks hinders the quantification of the efficiency of electron acceleration under realistic conditions.

The Earth's bow shock is a natural laboratory for probing the microphysics of supercritical collisionless shocks, because the Alfv\'{e}n Mach number of the solar wind flow typically exceeds the second critical value, $M_{A}\gtrsim 3$ \citep[e.g.,][]{Kennel85}. The {\it in-situ} measurements in the Earth's bow shock showed that electric and magnetic field fluctuations are electromagnetic below a few hundred Hz and mostly electrostatic at higher frequencies \citep[][]{Rodriguez75,Mozer13}. The measurements of electric and magnetic field waveforms demonstrated that the electromagnetic fluctuations correspond to whistler waves \citep[e.g.,][]{Wilson14,Oka17:apjl}, while the electrostatic turbulence corresponds to ion-acoustic waves \citep[][]{Balikhin05,Hull06,Goodrich18:iaw} and bipolar electrostatic structures \citep{Bale98,Bale02}. The bipolar structures were interpreted in terms of electron phase space holes, as electrostatic structures produced in a nonlinear stage of various electron streaming instabilities \citep[e.g.,][]{Schamel86}, and involved in the original scenario of electron surfing acceleration in high Mach number shocks \citep[][]{Hoshino02,Schmitz02b}. However, until recently, spacecraft measurements did not allow the resolution of the nature and generation mechanisms of the bipolar structures in the Earth's bow shock.

The recently launched Magnetospheric Multiscale (MMS) spacecraft \citep{Burch16} has allowed us to probe the Earth's bow shock with unprecedented temporal resolution and 3D electric field measurements. The analysis of about twenty bipolar structures measured in a particular Earth's bow shock crossing showed that these structures are not electron phase space holes because they have negative electrostatic potentials \citep{Vasko18:grl}. In this Letter, we present a statistical analysis of more than one hundred bipolar structures measured in the shock transition region of a particular Earth's bow shock crossing. We argue that the bipolar structures are ion phase space holes produced by the two-stream instability between incoming and reflected ions in the shock transition region. The implications for the electron surfing acceleration in collisionless shocks are discussed.

\begin{figure*}
\centering
\includegraphics[width=40pc]{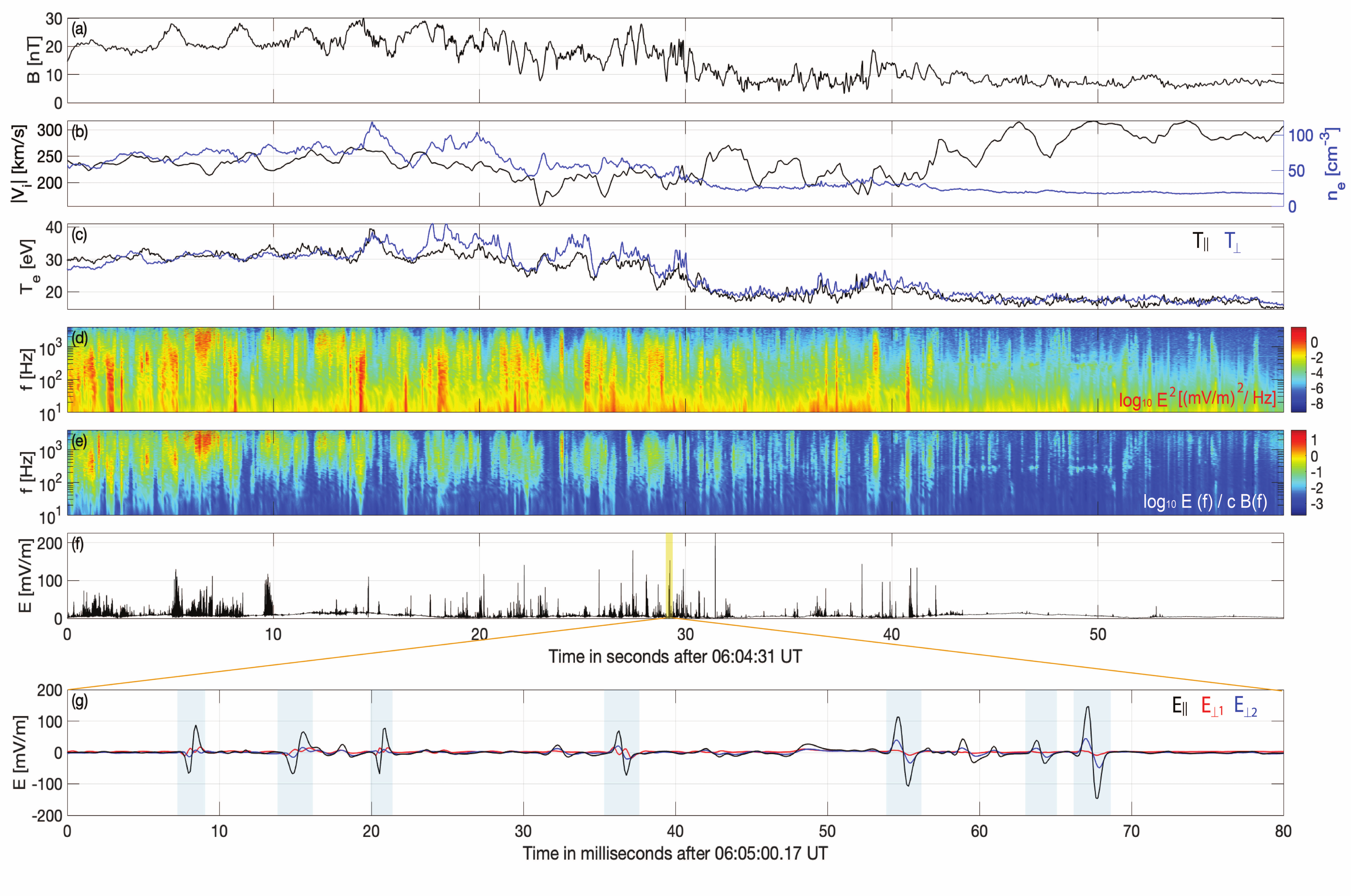}
\caption{Overview of the Earth's bow shock crossing by the Magnetospheric Multiscale spacecraft on November 2, 2017. The panels present measurements of MMS4, while other three spacecrafts, being located withing a few tens of kilometers, provide essentially identical overviews: (a) the magnitude of a quasi-static magnetic field measured at 128 samples/s; (b) the electron density (0.03s cadence) and the magnitude of the ion bulk velocity (0.15s cadence); (c) electron temperatures (0.03 cadence) parallel and perpendicular to a local quasi-static magnetic field; (d) total power spectral density (PSD) $E^2(f)$ of the electric field fluctuations (8,192 sample/s) computed using 0.1s sliding window, where $f$ denotes frequency (similar PSD $B^2(f)$ was computed for the magnetic field fluctuations measured at 8,192 samples/s); (e) the ratio $E(f)/cB(f)$ between PSDs of electric and magnetic field fluctuations ($c$ is the speed of light), where higher values above a few hundred Hz indicate that the electric field fluctuations at those frequencies tend to be electrostatic \citep[in accordance with][]{Rodriguez75}; (f) the amplitude of the electric field fluctuations measured at 8,192 samples/s; (g) an expanded view of three electric field components measured over 0.08s interval highlighted in panel (f), where $E_{||}$ is the electric field component parallel to a local quasi-static magnetic field, while $E_{\perp1}$ and $E_{\perp2}$ are corresponding perpendicular components. \label{fig1}}
\end{figure*}

\section{Observations}

We consider the Earth's bow shock crossing by the four MMS spacecrafts on November 2, 2017 around 06:03:00 UT. We use the DC-coupled magnetic
field (128 samples per second) provided by Digital and Analogue Fluxgate Magnetometers \citep{Russell16}, AC-coupled electric fields (8,192 samples per second) provided by Axial Double Probe \citep{Ergun16} and Spin-Plane Double Probe \citep{Lindqvist16}, AC-coupled magnetic fields (8,192 samples per second) provided by the Search Coil magnetometer\citep{LeContel16}, electron moments (0.03s cadence) and ion moments (0.15s cadence) provided by the Fast Plasma Investigation instrument \citep{Pollock16}. The electric field is measured by four voltage-sensitive spherical probes on 60-m antennas in the spacecraft spin plane (almost in the ecliptic plane) along with two probes on roughly 15-m axial antennas along the spin axis (almost perpendicular to the ecliptic plane). The voltages of the opposing probes measured with respect to the spacecraft are used to estimate the direction of propagation, velocity and other parameters of bipolar electrostatic structures \citep[see][for methodology details]{Vasko18:grl}. We determine the normal to the shock in the GSE (Geocentric Solar Ecliptic) coordinate system with the $z$-axis perpendicular to the ecliptic plane, the $x$-axis pointing to the Sun and the $y$-axis completing the right-hand coordinate system.

\begin{figure*}
    \centering
   \includegraphics[width=40pc]{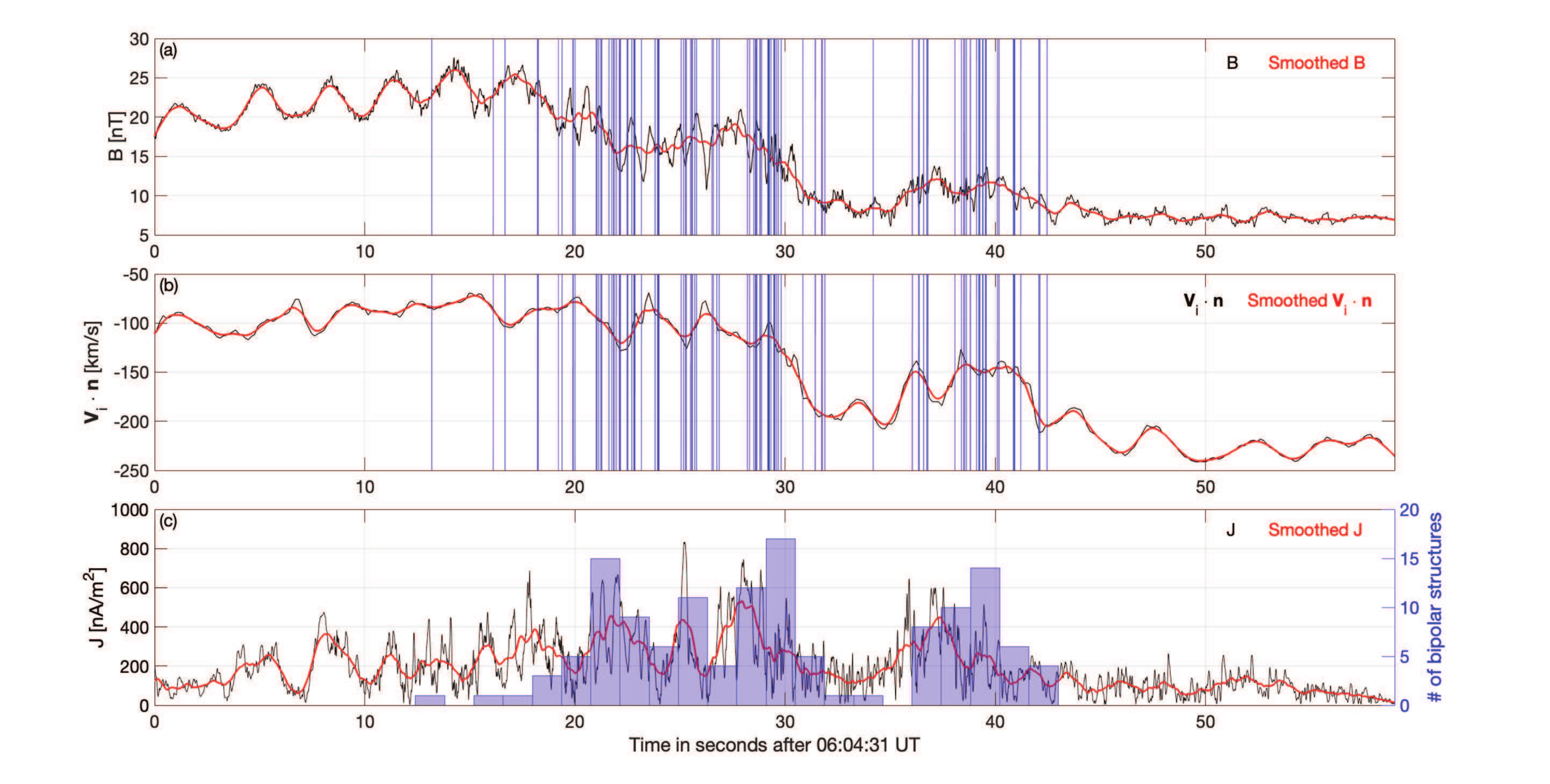}
    \caption{The analysis of occurrences of 134 bipolar electrostatic structures with amplitudes exceeding 50 mV/m that were selected using electric field measurements aboard four MMS spacecrafts: (a) the magnitude of the quasi-static magnetic field computed as an average value of the magnetic fields measured aboard four MMS spacecrafts (black) and its profile smoothed using 1.5s sliding window (red); the occurrence times (vertical lines) of the bipolar structures; (b) the ion bulk velocity (average value of ion bulk velocities measured aboard four MMS spacecraft) along the shock normal ${\bf n}$ (black) and its profile smoothed using 1.5s sliding window (red); the occurrence times (vertical lines) of the bipolar structures; (c) the magnitude of the current density computed using simultaneous magnetic field measurements aboard four MMS spacecrafts (black) and its profile smoothed using 1.5s sliding window (red); the histogram presents the number of bipolar structures observed within bins of 1.5s duration.
    \label{fig2}}
\end{figure*}

\begin{figure}[!t]
    \centering
    \includegraphics[width=20pc]{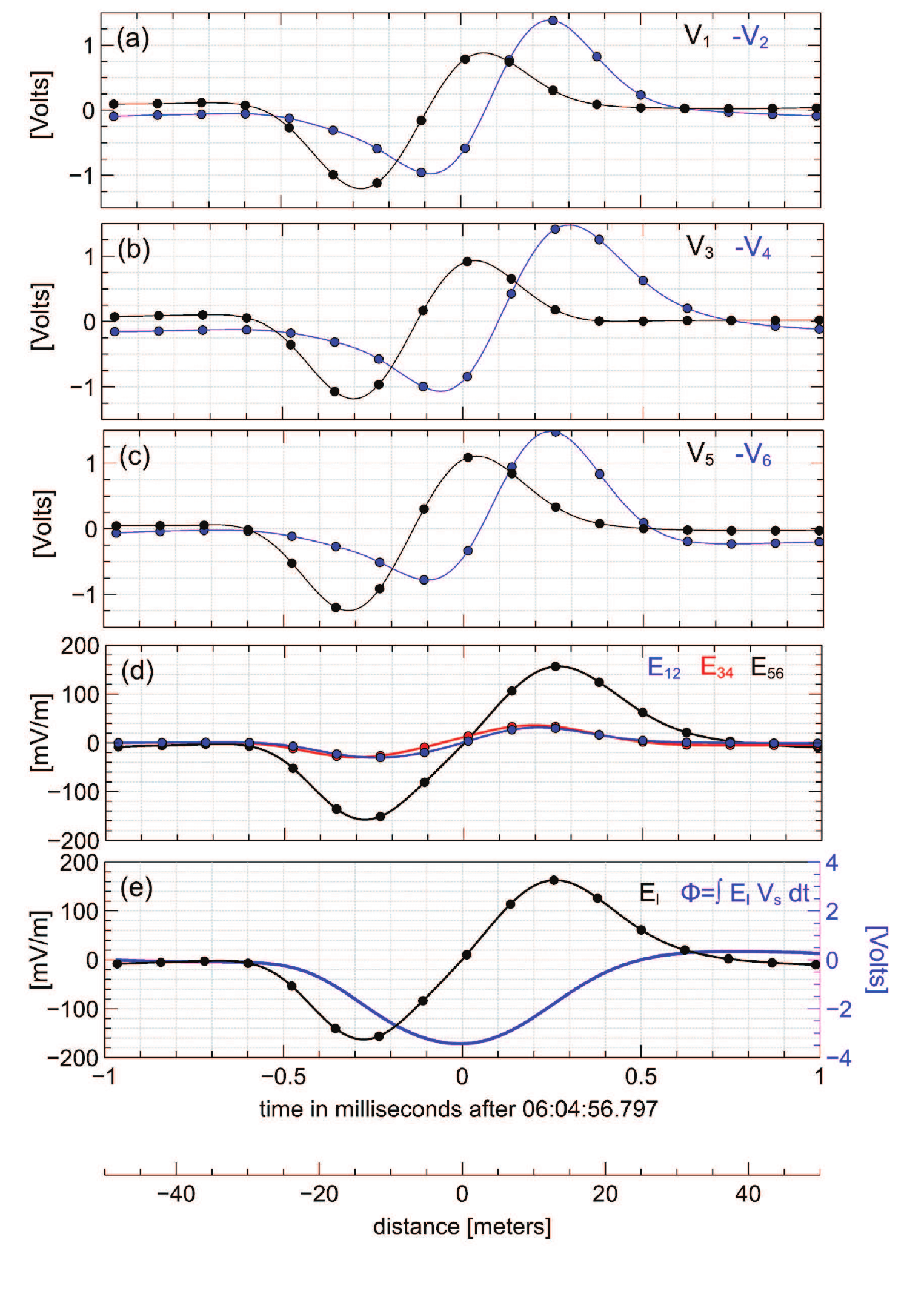}
    \caption{The analysis of properties of a particular bipolar structure measured aboard MMS4 that is based on voltage signals induced on six voltage-sensitive probes by the electric field of the bipolar structure \citep[see][for methodology details]{Vasko18:grl}: (a, b) voltage signals $V_{1}$ vs. $-V_{2}$ and $V_{3}$ vs. $-V_{4}$ of the opposing probes mounted on 60-m antennas in the spacecraft spin plane; (c) voltage signals $V_{5}$ vs. $-V_{6}$ of the opposing probes mounted on 15-m axial antennas along the spin axis; the time delays between voltage signals of the opposing probes are used to compute the direction of propagation ${\bf k}$ and velocity $V_{s}$ of the bipolar structure; (d) the electric field components $E_{12}$, $E_{34}$ and $E_{56}$ along the antenna directions that were computed using the voltage signals of the opposing probes, $E_{ij}\propto (V_{i}-V_{j})/(2l_{ij})$, where $l_{12}=l_{34}=60$ m and $l_{56}=15$ m are antenna lengths; (e) the electric field $E_{l}$ of the bipolar structure (black) oriented a few degrees off the axial antenna (as one can infer from similar bipolar profiles in panel (d)); the electrostatic potential of the bipolar structure (blue) is computed as $\Phi=\int {\bf E}\cdot {\bf k}\; V_{s}\;dt$. In all panels dots represent measured quantities, while solid lines correspond to spline interpolated quantities. The electrostatic potential $\Phi$ is computed using the interpolated $E_{l}$ profile. The lowest horizontal axis provides the spatial distance along the propagation direction ${\bf k}$ computed as $\int V_{s}dt$ and measured from $E_{l}=0$.\label{fig3}}
\end{figure}

Figure \ref{fig1} presents a summary of the Earth's bow shock crossing as measured aboard MMS4. The other MMS spacecraft being located within about twenty kilometers of MMS4 provide almost identical overviews of the shock. The shock transition region can be seen in panel (a) by the magnetic field increase from about 7 nT in the upstream region to about 20 nT in the downstream region. There is an associated deceleration of incoming solar wind ions and increase of the plasma density from the upstream value of 16 cm$^{-3}$ to the downstream value of 60 cm$^{-3}$ as shown in panel (b). The electron heating in the shock transition region is essentially isotropic, that is, parallel and perpendicular electron temperatures are almost identical as shown in panel (c). The electron temperature increases from about 15 eV in the upstream region to about 30 eV in the downstream region. The ion temperature in the upstream region is not well measurable by MMS, while Wind spacecraft\footnote{The website  \texttt{https://cdaweb.gsfc.nasa.gov/} provides Wind measurements of plasma parameters time-shifted to the nose of the Earth's bow shock.} provides an estimate of 6 eV.

The upstream and downstream values of the quantities presented in panels (a) and (b) are used for estimating the normal to the shock and velocity of the shock in the spacecraft frame using the Rankine-Hugoniot conditions \citep[][]{Vinas&Scudder86}. We have found that in the GSE coordinate system the normal to the shock is ${\bf n}\approx (0.81,0.56,0.2)$ and the shock propagates with the velocity of 38 km/s in the direction opposite to the normal, that is, toward the Earth. The shock is quasi-perpendicular where the angle between the normal and the upstream magnetic field is $\theta_{Bn}\approx 96^{\circ}$. In the rest frame of the shock, the ion bulk velocity along the normal decreases from about 200 km/s in the upstream region to about 70 km/s in the downstream region (not shown here). The upstream velocity of 200 km/s corresponds to the Alfv\'en Mach number $M_{A}\approx 5.4$. Thus, the considered shock is a supercritical quasi-perpendicular shock with $T_{e}/T_{i}\approx 2.5$ and $\beta_{i}=8\pi n T_i/B^2\approx 0.8$ in the upstream region. In this regime the magnetic field in the shock transition region is rather turbulent in accordance with numerical simulations \citep[e.g.,][]{Leroy82,Scholer03}.

\begin{figure*}
    \centering
    \includegraphics[width=40pc]{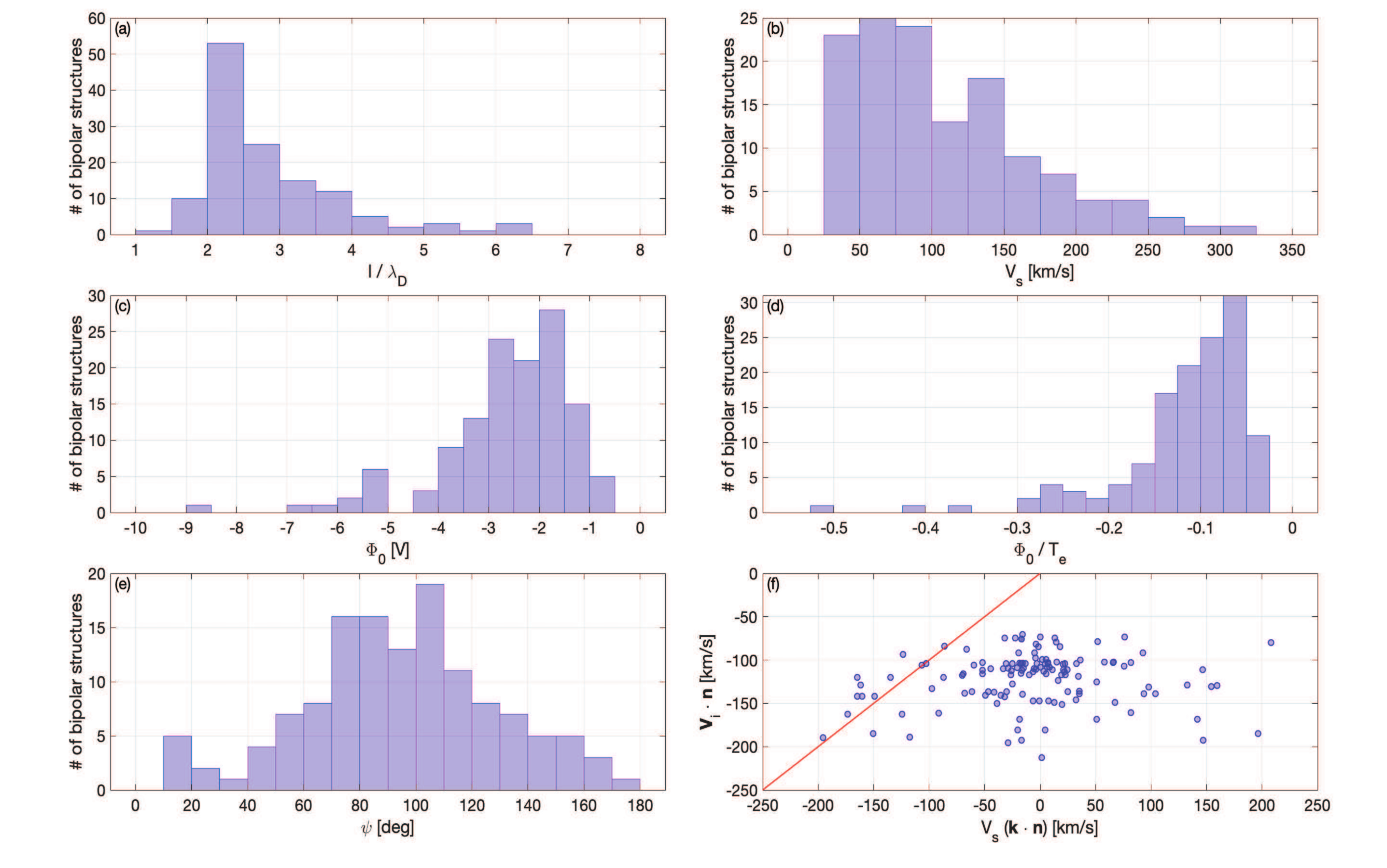}
    \caption{A summary of properties of 134 bipolar structures: (a) the spatial scale $l$ in units of local Debye length $\lambda_{D}$; (b) the velocity $V_{s}$ of the bipolar structures in the spacecraft reference frame; (c,d) the amplitude $\Phi_0$ of the electrostatic potential in physical units and in units of local electron temperature $T_{e}$; (e) the distribution of $\psi={\rm \cos}^{-1} ({\bf k}\cdot {\bf n})$, which is the angle between the propagation direction ${\bf k}$ of a bipolar structure and the shock normal ${\bf n}$; (f) the velocity of the bipolar structures along the normal, $V_{s} ({\bf k}\cdot {\bf n})$, versus the normal component of the local ion bulk velocity, ${\bf V}_{i}\cdot {\bf n}$; the red line corresponds to $V_{s} ({\bf k}\cdot {\bf n})={\bf V}_{i}\cdot {\bf n}$. For practically all bipolar structures we observed $V_{s} ({\bf k}\cdot {\bf n})>{\bf V}_{i}\cdot {\bf n}$, which means that in the plasma rest frame the bipolar structures propagate toward the upstream region.\label{fig4}}
\end{figure*}

We have computed power spectral densities (PSD) of electric and magnetic field fluctuations (8,192 samples/s) using 0.1s sliding window. The electric field PSD shown in panel (d) demonstrates the presence of broadband electric field fluctuations in the shock transition and downstream regions. The ratio between the electric and magnetic field PSD shown in panel (e) indicates that the electric field fluctuations above a few hundred Hz tend to be electrostatic in accordance with previous measurements \citep[][]{Rodriguez75}. Panel (f) shows that the electric field fluctuations in the shock transition region have amplitudes up to a few hundred mV/m. An expanded view of three electric field components measured over a highlighted 0.08s interval demonstrates that some of the intense electric field fluctuations are due to bipolar electrostatic structures with duration of a few milliseconds. A careful inspection through the electric field fluctuations with amplitudes exceeding 50 mV/m has resulted in a dataset of 134 bipolar structures observed aboard four MMS spacecrafts. In what follows we focus on analysis of these large-amplitude bipolar structures.

Figure \ref{fig2} presents analysis of the occurrence of the bipolar structures. Panel (a) shows that the bipolar structures occur predominantly in the shock transition region, and only a few bipolar structures are observed in the downstream region. In addition, the bipolar structures preferentially occur around the magnetic field gradients. Panel (b), which presents the ion bulk velocity along the shock normal, demonstrates that the magnetic field gradients are associated with the slowing down of the ion bulk flow. Panel (c) presents the distribution of the bipolar structures that is obtained by counting the number of bipolar structures within bins of 1.5s duration. In addition, panel (c) presents the magnitude of a local current density estimated using simultaneous magnetic field measurements aboard four MMS spacecrafts \citep[see, e.g.,][for methodology]{Chanteur98:issi} along with its profile smoothed using 1.5s sliding window. The occurrence of the bipolar structures is well seen to be correlated with the local current density magnitude which is equivalent to the correlation with the magnetic field gradients in the shock transition region. This feature of the occurrence of bipolar structures in collisionless shocks is reported for the first time and will be discussed in the next section.

Figure \ref{fig3} presents analysis of properties of a particular bipolar structure measured aboard MMS4. The analysis is based on voltage signals induced on voltage-sensitive probes by the electric field of the bipolar structure \citep[see][for methodology details]{Vasko18:grl}. Panels (a) and (b) present voltage signals measured by two pairs of opposing probes on 60-m antennas in the spacecraft spin plane, while panel (c) presents voltage signals measured by the two opposing probes on 15-m axial antennas along the spin axis. Panel (d) presents components of the electric field ${\bf E}$ along the antenna directions computed using the voltage signals of the opposing probes. The time delays between the voltage signals of the opposing probes well noticeable in panels (a)-(c) allow the estimation of velocity and direction of propagation of the bipolar structure. We have found that the bipolar structure propagates with velocity $V_{s}\approx 62$ km/s along a unit vector ${\bf k}$ that is just a few degrees off the axial antenna. Interestingly, the bipolar structure propagates highly oblique to the shock normal, $\psi={\rm cos}^{-1} ({\bf k}\cdot {\bf n})\approx 90^{\circ}$. Panel (d) shows that all three electric field components have similar bipolar profiles, while the electric field along the axial antenna is the dominant component. This indicates that the electric field of the bipolar structure is oriented a few degrees off the axial antenna direction. Panel (e) presents the electric field $E_l$ in that direction, while the other two components are negligible compared to $E_{l}$ (not shown here). Because both ${\bf k}$ and ${\bf E}$ are approximately along the axial antenna, the angle between them is just a few degrees, indicating that the bipolar structure is approximately a 1D structure.

The estimated velocity of the bipolar structure allows the translation of temporal profiles into spatial profiles with a spatial coordinate along the propagation direction ${\bf k}$. The spatial coordinate measured from $E_l=0$ is given below panel (e). We have computed the electrostatic potential of the bipolar structure as $\Phi=\int {\bf E}\cdot {\bf k}\;V_{s}\;dt$. Panel (e) shows that the bipolar structure has a negative electrostatic potential with a peak value $\Phi_0\approx -3.5$ V or $\Phi_0\approx -0.1\;T_{e}$ in units of local electron temperature. We define the spatial scale $l$ of the bipolar structure as $l=0.5\;V_{s}\Delta t$, were $\Delta t$ is the time interval between minimum and maximum values of $E_l$. Panel (e) shows that the spatial scale of the bipolar structure is $l\approx 16$ m or $l\approx 2\lambda_{D}$ in units of local Debye lengths. We have performed similar analysis of properties of all 134 bipolar structures and found that all of the bipolar structures have negative electrostatic potentials and hence cannot be interpreted in terms of electron phase space holes \citep[e.g.,][]{Schamel86}. We have also found that for more than $80$\% of the bipolar structures, the angle between ${\bf k}$ and ${\bf E}$ is within 30$^{\circ}$, so most of the bipolar structures are approximately 1D structures.

Figure \ref{fig4} presents statistical distributions of the estimated parameters of the bipolar structures. Panel (a) shows that the bipolar structures have typical spatial scales of a few local Debye lengths that is less than one tenth of electron thermal gyroradius (not shown here). Panel (b) shows that bipolar structures commonly propagate with velocity around 100 km/s and higher velocities are rarer. Panels (c) and (d) show that the amplitudes of the electrostatic potential of the bipolar structures are typically a few Volts and within a few tenths of a local electron temperature. Panel (e) presents the distribution of $\psi={\rm cos}^{-1} ({\bf k}\cdot {\bf n})$, which indicates that the bipolar structures propagate highly oblique to the shock normal: $45^{\circ}\lesssim \psi\lesssim 135^{\circ}$ for more than 80$\%$ of the structures and $60^{\circ}\lesssim \psi\lesssim 120^{\circ}$ for more than 65\% of the structures. Panel (f) presents a comparison between $V_{s}({\bf k}\cdot{\bf n})$, the velocity of bipolar structures along the shock normal, and ${\bf V}_{i}\cdot {\bf n}$, the ion bulk velocity component along the shock normal (see also Figure \ref{fig2}b). In the spacecraft frame the plasma flows toward the downstream region, ${\bf V}_{i}\cdot{\bf n}<0$, while the bipolar structures can propagate both toward the upstream, ${\bf k}\cdot{\bf n}>0$, and downstream, ${\bf k}\cdot {\bf n}<0$ regions. Interestingly, in the plasma rest frame, practically all bipolar structures propagate toward the upstream region, because as shown in panel (f) we observe $V_{s}({\bf k}\cdot{\bf n})>{\bf V}_{i}\cdot{\bf n}$ for all bipolar structures, except for several structures satisfying $V_{s}\;({\bf k}\cdot{\bf n})\approx {\bf V}_{i}\cdot{\bf n}$. This feature of propagation direction of the bipolar structures is reported for the first time and will be discussed in the next section.

\section{Interpretation}

We have demonstrated that the large-amplitude bipolar structures observed in the shock transition region are Debye-scale structures with a negative electrostatic potential, propagating highly oblique to the shock normal. In the plasma rest frame, bipolar structures propagate toward the upstream region. The occurrence of bipolar structures is tightly correlated with magnetic field gradients in the shock transition region. These properties reveal the nature of the bipolar structures and instability driving them in the shock transition region.

The negative electrostatic potential of bipolar structures leads to the interpretation of these structures in terms of ion phase space holes, which are electrostatic structures formed in a nonlinear stage of various ion streaming instabilities \citep[e.g.,][]{Schamel86,Kofoed-Hansen89,Borve01}. Ion phase space holes are formed from ions trapped in potential wells of electrostatic fluctuations driven by instability. Regardless of the instability that produces bipolar structures in the shock transition region, there is a lowest increment value for that instability to be capable of producing the observed bipolar structures. Because the instability saturation occurs, when the bounce period of ions trapped within electrostatic fluctuations becomes comparable to an initial increment \cite[e.g.,][]{Sag&Gal69}, that increment $\gamma$ should exceed the bounce frequency of ions trapped within bipolar structures, $\omega_{b}\approx l^{-1} (e|\Phi_0|/m_{i})^{1/2}$, where $m_{i}$ is the ion mass, $l$ and $\Phi_0$ are the spatial scale and amplitude of the electrostatic potential of a bipolar structure respectively. We rewrite the criterion $\gamma\gtrsim \omega_{b}$ as follows
\begin{eqnarray}
\frac{\gamma}{\omega_{pi}}\gtrsim  \frac{\lambda_{D}}{l}\left(\frac{e|\Phi_0|}{T_{e}}\right)^{1/2}
\label{eq:gamma_req}
\end{eqnarray}
where $\omega_{pi}=(4\pi n_0 e^2/m_{i})^{1/2}$ is the ion plasma frequency. Adopting typical parameters of the observed bipolar structures, $l/\lambda_{D}\sim 2$ and $e|\Phi_0|/T_{e}\sim 0.1$, we find that the initial increment should be of the order of a fraction of the ion plasma frequency, $\gamma\sim 0.1\;\omega_{pi}$.

\begin{figure}[!t]
    \centering
    \includegraphics[width=20pc]{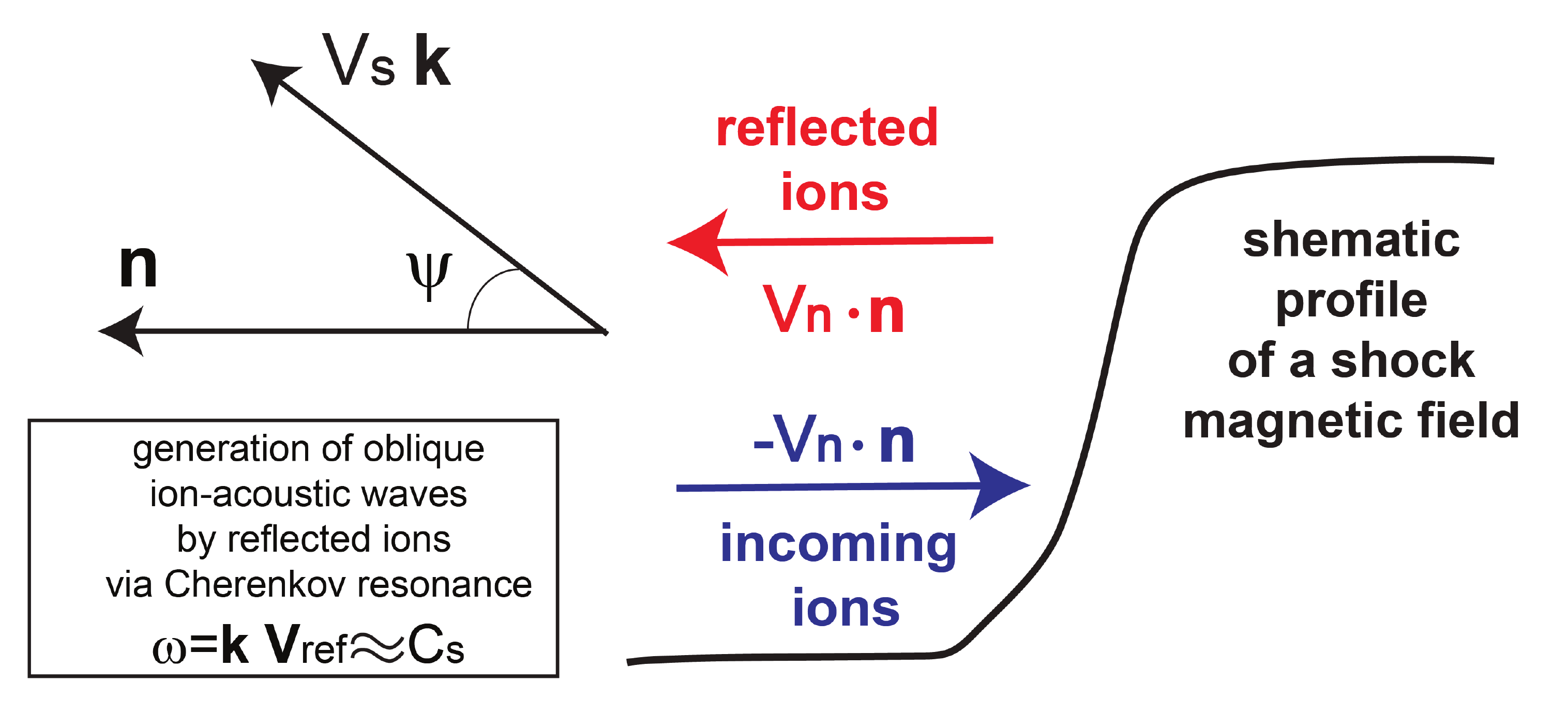}
    \caption{A schematic of the ion two-stream instability between incoming and reflected ions in the shock transition region. In the normal incidence frame the bulk velocity of incoming ions is $-{V}_{n} {\bf n}$, where $V_{n}=|{\bf V}_{i}\cdot{\bf n}-V_{sh}|$, $V_{sh}$ is the shock velocity and ${\bf n}$ is the shock normal. In the frame of incoming ions the reflected ions propagate upstream with velocity ${\bf V}_{\rm ref}=2 V_{n} {\bf n}$. The reflected ions are capable of driving ion-acoustic waves satisfying the Cherenkov resonance, $\omega={\bf k}\cdot {\bf V}_{\rm ref}$, where frequency $\omega$ and wave vector ${\bf k}$ are related to each other by the ion-acoustic wave dispersion relation. In the rest frame of incoming ions the ion-acoustic waves propagate toward the upstream region, have wavelengths of a few Debye lengths and propagate oblique to the shock normal at an angle $\psi$ satisfying $\cos\psi\approx c_{s}/V_{\rm ref}$, where $c_{s}$ is the ion-acoustic velocity. In a nonlinear stage of the instability the ion-acoustic waves transform into ion phase space holes \citep[e.g.,][for simulations]{Kofoed-Hansen89,Borve01}. The ion phase space holes inherit the properties of the ion-acoustic waves: propagate in the direction of reflected ions, which is toward the upstream region (in the rest frame of incoming ions), have wavelengths of the order of a few Debye lengths, and propagate highly oblique to the shock normal. \label{fig5}}
\end{figure}

The most plausible instability driving the observed bipolar structures is the ion two-stream instability between incoming and reflected ions \citep[e.g.,][]{Akimoto85,Ohira08}. First, the observed strong correlation between occurrence of the bipolar structures and magnetic field gradients indicates that reflected ions might be a source of free energy for the bipolar structures, because the reflection of a fraction of incoming ions is expected to occur due to magnetic field gradients \citep[e.g.,][]{Leroy82}. The observed deceleration of the ion bulk flow associated with the magnetic field gradients is due to that reflection of incoming ions (Figure \ref{fig2}). Second, the ion two-stream instability is capable of explaining the observed properties of the bipolar structures and capable of providing the required linear increments.

Figure \ref{fig5} presents a schematic of the ion two-stream instability in the shock transition region. Due to reflection of a fraction of incoming ions by a magnetic field gradient, the ion distribution function is locally a combination of incoming ions with density $n_0$ and reflected ions with density $n_{\rm ref}$. In the normal incidence frame the bulk velocity of incoming ions is $-V_{n}{\bf n}$, where $V_{n}=|{\bf V}_{i}\cdot {\bf n}-V_{sh}|$ and $V_{sh}$ is the shock velocity and ${\bf n}$ the shock normal. In the reference frame of incoming ions, reflected ions propagate along the shock normal (toward upstream) with velocity ${\bf V}_{\rm ref}=V_{\rm ref}{\bf n}=2V_{n}{\bf n}$. The simplest analysis of the instability between incoming and reflected ions was presented by \cite{Akimoto85} and \cite{Ohira08} by assuming cold ion populations and neglecting effects of the magnetic field (that is reasonable for waves with wavelengths much smaller than electron and ion thermal gyroradii, which is the case for Debye-scale waves). That analysis showed that reflected ions drive ion-acoustic waves satisfying the Cherenkov resonance
\begin{eqnarray}
\omega\approx {\bf k} {\bf V}_{\rm ref}=V_{\rm ref}({\bf k}\cdot {\bf n})=kV_{\rm ref}\cos\psi,
\label{eq:Cherenkov}
\end{eqnarray}
where $\psi$ is the angle between ${\bf k}$ and ${\bf n}$, and frequency $\omega$ and wave vector ${\bf k}$ are approximately related by the dispersion relation of ion-acoustic waves
\begin{eqnarray}
\omega\approx \omega_{pi} k\lambda_{D}/(1+{ k}^2\lambda_{D}^2)^{1/2}
\label{eq:iaw}
\end{eqnarray}
The fastest growing ion-acoustic waves have wavelengths of a few Debye lengths, $k\lambda_{D}\sim 1$, and the increment dependent on the fraction of reflected ions
\begin{eqnarray}
\frac{\gamma_{\rm max}}{\omega_{pi}}\approx \left(\frac{3\sqrt{3}}{16}
\frac{n_{\rm ref}}{n_0}\right)^{1/3}
\label{eq:gamma_max}
\end{eqnarray}
The resonance condition $\omega\approx kV_{\rm ref}\cos\psi$ shows that the fastest-growing ion-acoustic waves propagate oblique to the shock normal
\begin{eqnarray}
\cos\psi\approx \omega/kV_{\rm ref}\approx c_{s}/V_{\rm ref},
\label{eq:psi}
\end{eqnarray}
where $c_{s}=\omega_{pi}\lambda_{D}=(T_{e}/m_{i})^{1/2}$ is the ion-acoustic velocity. Thus, ion-acoustic waves produced by the instability between incoming and reflected ions: {\bf (1)} propagate in the direction of reflected ions, that is, toward the upstream region (in the rest frame of incoming ions); {\bf (2)} have wavelengths of the order of a few Debye lengths; {\bf (3)} propagate oblique to the shock normal.

The properties (1)-(3) above are consistent with the observed parameters of the bipolar structures. We have found that the bipolar structures propagate toward the upstream region in the plasma rest frame. In that frame, the incoming ions propagate toward the downstream region, while reflected ions propagate upstream. Therefore, in the rest frame of incoming ions, the bipolar structures also propagate toward the upstream region that is in accordance with (1). The bipolar structures have spatial scales of a few Debye lengths and propagate oblique to the shock normal that is in accordance with (2) and (3). The observed highly oblique propagation results from the Cherenkov resonance condition, $\cos\psi\approx c_{s}/V_{\rm ref}$, where $c_{s}=(T_{e}/m_{i})^{1/2}$ is of the order of 50-100 km/s, $V_{\rm ref}=2\;|{\bf V_{i}}\cdot {\bf n}-V_{sh}|$ is in the range from 400 to 120 km/s, because $V_{sh}\approx -38$ km/s and ${\bf V}_{i}\cdot {\bf n}$ is in the range from -250 to -100 km/s (Figure \ref{fig2}). Finally, according to Eq. (\ref{eq:gamma_max}) for typical densities of reflected ions, $n_{\rm ref}\sim 0.1\;n_0$ \citep{Leroy82,Scholer03}, the ion two-stream instability can provide initial increments of a fraction of the ion plasma frequency as required by Eq. (\ref{eq:gamma_req}).

We have assumed both incoming and reflected ions to be cold. Finite ion temperatures would affect the instability characteristics quantitatively, but not the most critical features of the ion two-stream instability \citep[][]{Gary87}: propagation in the same direction as reflected ions (in the rest frame of incoming ions), wavelengths of a few Debye lengths, and highly oblique propagation to the shock normal. Therefore, we consider our interpretation to be robust.


\section{Discussion}

The bipolar structures in the Earth's bow shock were originally interpreted in terms of electron phase space holes, which are electrostatic structures produced in a nonlinear stage of various electron streaming instabilities \citep[][]{Bale98,Bale02}. The potential instabilities were electron two-stream \citep[e.g.,][]{Gedalin99} and beam \citep[e.g.,][]{Thomsen83} instabilities. However, the recent analysis of about twenty bipolar structures in a particular Earth's bow shock crossing showed that the bipolar structures cannot be electron phase space holes, because they have a negative electrostatic potential \citep{Vasko18:grl}. In this Letter we have considered an Earth's bow shock crossing with more than one hundred bipolar structures in the shock transition region and confirmed that the bipolar structures cannot be electron phase space holes. Based on the detailed analysis, we have interpreted the bipolar structures in terms of ion phase space holes produced by the instability between incoming and reflected ions. That is the first experimental evidence that the ion two-stream instability produces the electrostatic turbulence in collisionless shocks.

The ion two-stream instability between incoming and reflected ions was suggested by \cite{Akimoto85}, while \cite{Ohira08} have recently revived interest to that instability. The 2D Particle-In-Cell (PIC) simulations of the ion two-stream instability evolution in a uniform plasma have demonstrated ion heating and practically no electron heating or acceleration \citep{Ohira08}. However, as discussed below, we cannot rule out that in a realistic non-uniform shock configuration, the electrostatic turbulence driven by the ion two-stream instability is capable of accelerating a fraction of thermal electrons to superthermal energies.

The 2D PIC simulations by \cite{ohira07} showed that in a uniform plasma the electrostatic turbulence driven by the Buneman instability (typical of high Mach number shocks) is incapable of accelerating electrons via the surfing mechanism demonstrated by 1D simulations \citep{Hoshino02}. On the contrary, the 2D PIC simulations of \cite{Amano09}, which included a realistic non-uniform shock configuration, demonstrated that the Buneman instability can provide electron acceleration via stochastic surfing acceleration (SSA) mechanism. In the SSA mechanism electrons are accelerated to superthermal energies due to multiple interactions with the electrostatic turbulence in the upstream region, which are possible due to electron mirroring by a non-uniform magnetic field of the shock.

The recent 2D PIC simulations by \cite{Umeda09} have demonstrated that the SSA mechanism can also operate at low Mach numbers typical of the Earth's bow shock. In those simulations the electrostatic turbulence is produced by reflected ions. Although \cite{Umeda09} did not dwell into the nature of the instability, the most plausible case is the ion two-stream instability. The identification of the ion two-stream instability presented in this Letter and simulations by \cite{Umeda09} indicate that the electrostatic turbulence produced by that instability can provide electron acceleration in collisionless shocks via the SSA mechanism.

\section{Conclusion}

The analysis of more than one hundred bipolar structures in a supercritical quasi-perpendicular Earth's bow shock showed that the bipolar structures are ion phase space holes produced by the two-stream instability between incoming and reflected ions. The arguments supporting this interpretation are
\begin{enumerate}
\item the bipolar structures have negative amplitudes of the electrostatic potential and spatial scales of a few Debye lengths.

\item the occurrence of the bipolar structures is correlated with the magnetic field gradients capable of reflecting a fraction of incoming ions.

\item in the shock rest frame the bipolar structures propagate highly oblique to the shock normal, the angle between the propagation direction and the shock normal is within (45$^{\circ}$, 135$^{\circ}$) for more than $80\%$ of the bipolar structures.

\item in the plasma rest frame the bipolar structures propagate toward the upstream region, that is, in the direction of propagation of reflected ions.

\item the ion two-stream instability is capable of providing the required increments of a fraction of the ion plasma frequency.
\end{enumerate}
That is the first demonstration that the ion two-stream instability produces the electrostatic turbulence in supercritical collisionless shocks.

\acknowledgments
The work was supported by NASA MMS Guest Investigator grant No. 80NSSC18K0155. I.V. also thanks for support the International Space Science Institute, Bern, Switzerland. A.A. thanks Russian Science Foundation for support through grant No. 19-12-00313. We thank the MMS teams for the excellent data. The data are publicly available at https://lasp.colorado.edu/mms/public.

\end{document}